\documentstyle[graphics,twocolumn]{mn}
\newcommand{\mincir}{\raise
-2.truept\hbox{\rlap{\hbox{$\sim$}}\raise5.truept 
\hbox{$<$}\ }}
\newcommand{\magcir}{\raise
-2.truept\hbox{\rlap{\hbox{$\sim$}}\raise5.truept
\hbox{$>$}\ }}
\newcommand{\minmag}{\raise-2.truept\hbox{\rlap{\hbox{$<$}}\raise
6.truept\hbox
{$>$}\ }}
\newcommand{\bfx}{{\bf x}}
\renewcommand{\H}{{\mbox{${\rm H{\sc i}~}$}}}

\newcommand{\eg}{{\it e.g.~}}
\newcommand{\ie}{{\it i.e.~}}
\newcommand{\mpc}{\rm Mpc}
\newcommand{\be}{\begin{equation}}
\newcommand{\ee}{\end{equation}}
\newcommand{\ba}{\begin{eqnarray}}
\newcommand{\ea}{\end{eqnarray}}
\newcommand{\brr}{\begin{array}}
 
\newcommand{\err}{\end{array}}
\newcommand{\bc}{\begin{center}}
\newcommand{\ec}{\end{center}}

\newcommand{\vel}{\,{\rm km\,s^{-1}}}

\title[ Modelling of the gas distribution: the Ly$\alpha$ 
forest] {Modelling the IGM and the Ly$\alpha$ forest at 
high redshift from the dark matter distribution}
\author[M. Viel, S. Matarrese, H. J. Mo, T. Theuns \& M. G. Haehnelt] 
{M. Viel $^{1, 2, 3}$,
S. Matarrese $^1$, H. J. Mo $^2$, Tom Theuns $^{3}$ \&
M. G. Haehnelt $^3$ \\ $^1$ Dipartimento di Fisica `Galileo Galilei',
via Marzolo 8, I-35131 Padova, Italy \\ $^2$ Max-Planck-Institut f\"ur
Astrophysik, Karl-Schwarzschild-Strasse 1, D-85741 Garching, Germany\\
$^3$ Institute of Astronomy, Madingley Road, Cambridge CB3 0HA, UK.\\
Email: viel@pd.infn.it, matarrese@pd.infn.it, hom@mpa-garching.mpg.de,
tt@ast.cam.ac.uk., haehnelt@ast.cam.ac.uk}

\date{\bf MNRAS accepted}

\begin{document}

\maketitle

\begin{abstract}
A variety of approximate schemes for modelling the 
low-density  Intergalactic Medium (IGM) in the high-redshift Universe 
is compared to the results of a large high-resolution hydro-dynamical 
simulation. These  schemes use either an analytical description of 
the dark matter distribution and the 
IGM  or numerical simulations of the DM distributions
combined with different approximate relations between dark matter field and 
the gas distribution.  Schemes based on a filtering of the 
dark matter distribution with a global Jeans scale result in a rather poor 
description of the gas distribution. An adaptive filtering which takes into 
account the density/temperature  dependence of the Jeans scale is required. A
reasonable description of the gas distribution can be achieved using a 
fit of the mean relation between the dark matter and gas  densities in
the  hydro-dynamical simulation to relate dark matter and gas 
distribution. In the hydro-dynamical simulations deviations from this mean 
relation are correlated with gradients in the dark matter peculiar velocity 
field  indicative of shocks in the gas component. A scheme which takes into 
account this correlation  results in a further improved gas distribution. 
Such adaptive filtering schemes  applied to  dark matter simulations will
be very well suited for studies of statistical properties of the Ly$\alpha$ 
forest  which investigate the IGM and the underlying dark matter distribution 
and require a large dynamic range and/or an extensive parameter study. 
\end{abstract}


\section{Introduction}

The low-density Intergalactic Medium (IGM) offers a unique and powerful
probe of the high redshift Universe.  Numerous weak absorption lines
seen in the spectra of distant quasars (the Ly$\alpha$ forest) are
produced by the small residue of neutral hydrogen in filamentary
structures intersected by the line of sight (Bahcall \& Salpeter 1965;
Gunn \& Peterson 1965; see Rauch 1998 for a recent review). Such
structures arise naturally in hierarchical cold dark matter dominated
models of structure formation, in which the IGM is highly ionised by
the UV-background produced by stars and galaxies. Simulations show that
the low-column density (${\rm N}_{\rm HI} \leq 10^{14.5}$ cm$^{-2}$)
absorption lines are produced by small fluctuations in the warm ($T\sim
10^{4}$ K) photo-heated IGM, which smoothly traces the mildly non-linear
dark matter filaments and sheets on scales larger the Jeans scale (Cen
{\it et al.} 1994, Petitjean {\it et al.} 1995, Miralda-Escud\'e {\it et al.}  1996, Zhang {\it et
al.}  1998).

This picture is supported by analytical studies based on simple models
for the IGM dynamics.  Such models are based on either a local
non-linear mapping of the linear density contrast, obtained for example by
applying a lognormal transformation (Coles \& Jones 1991) to the IGM
(Bi, B\"orner \& Chu 1992; Bi 1993; Bi, Ge \& Fang 1995; Bi \& Davidsen
1997, hereafter BD97), or on suitable modifications of the Zel'dovich
approximation (Zel'dovich 1970), to account for the smoothing caused by
the gas pressure on the Jeans scale (Reisenegger \& Miralda-Escud\'e 1995;
Gnedin \& Hui 1998; Hui, Gnedin \& Zhang 1997; Matarrese \& Mohayaee
2002). Many properties of the Ly$\alpha$ absorbers can be understood
with straightforward physical arguments (Schaye 2001, Zhang {\it et
al.} 1998).

The most convincing support for this picture comes, however, from
the comparison of observed spectra with mock spectra computed from
hydrodynamical numerical simulations (Cen {\it et al.} 1994; Zhang,
Anninos \& Norman 1995, 1997; Miralda-Escud\'e {\it et al.} 1996;
Hernquist {\it et al.}  1996; Charlton {\it et al.}  1997; Theuns {\it
et al.} 1998).  These simulated spectra accurately reproduce many observed
properties of the Ly$\alpha$ forest. Although there are still 
some discrepancies between observed and simulated spectra,  
especially in the Doppler parameters of the  absorption 
lines (Theuns {\it et al.} 1998; Bryan {\it et al.} 1999; Meiksin {\it et al.} 2001).

Observationally, the unprecedented high resolution observations of the
HIRES and UVES spectrographs on the Keck and VLT telescopes, 
as well as observations with the Hubble Space Telescope,
have lead to great advances in the understanding of the Ly$\alpha$
forest.  HIRES allowed to detect lines with column densities as low as
${\rm N}_{\rm HI} \sim 10^{12}$ cm$^{-2}$, while HST made a detailed
analysis of the low-redshift Ly$\alpha$ forest at $z<1.6$ possible.
Recent results include limits on the baryon density (Rauch {\it et
al.}  1997), the temperature and equation of state of the IGM (Schaye
{\it et al.}  2000; Theuns {\it et al.} 2001; Ricotti, Gnedin \&
Shull 2000; Bryan \& Machachek 2000, McDonald {\it et al.} 2001), the power
spectrum of the density fluctuations (Croft {\it et al.} 1998, 2001,
Gnedin \& Hamilton 2001, Zaldarriaga {\it et al.} 2001), the geometry
of the Universe (Hui {\it et al.} 1999; McDonald {\it et al.} 2001;
Viel {\it et al.} 2002) and direct inversions of the density field
(Nusser \& Haehnelt 1999, 2000; Pichon {\it et al.} 2001). Kim {\it et
al.}  (2001) used high resolution VLT/UVES data, to make an extensive
analysis of the Ly$\alpha$ forest in the redshift range $1.5 < z <
4$. Dobrzycki {\it et al.} (2001) presented results on the clustering
and evolution of the lines at $z<1.7$ using the HST/FOS spectrograph.

Many aspects of the warm photo-ionised Intergalactic medium can be well
modelled by hydrodynamical simulations. Hydro simulations are, 
however, still rather limited in dynamic range. The box size of 
hydro simulations which resolve the Jeans mass of the photo-ionised 
IGM with a temperature  of $\sim 10^4K$ probe the large scale
fluctuations of the density field rather poorly. This leads to 
uncertainties due to cosmic variance in the fluctuations on scales 
approaching the box size and missing fluctuations on scales larger 
than the simulation box. Because of limited computational resources 
it is also hardly possible to perform extensive parameter studies.  
In order to overcome these problems approximate methods for 
simulating the Ly$\alpha$ forest in QSO absorption spectra are 
often used (e.g. Gnedin \& Hui 1998; Croft {\it et al.} 1998; Croft
{\it et al.} 1999; Meiksin \& White 2001).  We test here 
a wide range of such approximate methods against a large
high-resolution hydro-dynamical simulation. 

The plan of the paper is as follows. In Section \ref{hydrosim} we
briefly describe the hydro-simulation used. 
Section \ref{secmet} presents
the lognormal model for the IGM and an improved model based
on the implementation of the gas probability distribution function taken from the
hydro-simulation. 
We improve our modelling in Section \ref{allmeth}, using 
two different ways of filtering the linear dark matter density field 
to model pressure effects. 
In this Section we present a further improvement, 
based on modelling the gas distribution starting from the non-linear 
dark matter density field obtained from the hydro-simulation and we show how
the smoothing on the Jeans length of the evolved dark-matter density
field poorly reproduces the real gas distribution.  In Section \ref{fluxstat}  
we compare the different methods proposed in terms of the 1-point and
2-points PDF of the 
flux. Section \ref{disc} contains a general
discussion and our conclusions.

\section{The hydro-dynamical simulation}
\label{hydrosim}
In the following Sections we will test our approximate 
schemes to model the Ly$\alpha$ forest against a large 
high-resolution hydro-dynamical simulation. We therefore 
summarise here some parameters of the hydro-dynamical simulation
used. The simulation techniques are described in more detail in 
Theuns {\it et al.} (1998).   We analyse a total of 7 outputs at 
redshifts $z=49$, $z=10$, $z=4$, $z=3.5$, $z=3$, $z=2.25$, $z=2$ of 
a periodic, cubic region in a $\Lambda$CDM Universe. The cosmological 
parameters are: $\Omega_{0m} = 0.3$, $\Omega_{0\Lambda} = 0.7$, 
$H_0=100h$ km s$^{-1}$Mpc with $h=0.65$ and
$\Omega_{0b} h^2=0.019$.  The comoving size of the box is $12.0/h$
Mpc. There are $256^3$ DM particles and $256^3$ gas particles, whose
masses are $m_{DM}=1.13\times 10^7 M_{\odot}$ and $m_{IGM}=1.91\times
10^6 M_{\odot}$. The input linear power spectrum was computed with
CMBFAST (Seljak \& Zaldarriaga 1996), and normalised to the abundance
of galaxy clusters using $\sigma_8=0.9$ (Eke, Cole \& Frenk 1996),
where $\sigma_8$ denotes the mass fluctuations in spheres of radius $8
h^{-1}$ Mpc.

The simulation code is based on HYDRA (Couchman {\it et al.} 1995) and 
combines Smoothed Particle Hydrodynamics (see {\it e.g.}  Monaghan 1992) with
P3M for self-gravity. Spline interpolation over gas particles allows
the computation of smooth estimates for density, temperature and
velocity and their gradients. The width of the spline kernel is matched
to the local particle number density, in this way high density regions
have higher numerical resolution than lower density
ones. Photo-ionisation and photo-heating rates are computed using the
fits in Theuns {\it et al.} (1998). 

Along many randomly chosen sight lines parallel to one of the axis of
the simulation box, we compute the gas and DM (over) density and
peculiar velocity, the gas temperature, and the neutral hydrogen
(density weighted) density, temperature and peculiar velocity. This
allows us to compute the absorption spectra. The smooth estimate for
the dark matter fields are computed using SPH interpolation as well,
with a smoothing length chosen to given of order $\sim 32$ neighbour
contributions per particle (Appendix A).

\section{Models using analytic descriptions of the dark 
matter distribution}
\label{secmet}

\subsection{The lognormal model}
\label{sub1}
We have started  with the model introduced by Bi and collaborators (Bi
{\it et al.} 1992, 1995; Bi 1993; BD97) for generating a Ly$\alpha$
absorption spectrum along a LOS. This simple model predicts many
properties of the absorption lines, including the column density
distribution and the distribution of line widths ($b$ parameters),
which can be directly compared with observations (BD97). Recently, the
BD97 model has been used by Roy Choudhury {\it et al.} (2000, 2001) to
study neutral hydrogen correlation functions along and transverse to
the line-of-sight. Feng \& Fang (2000) adopted the BD97 method to
analyse non-Gaussian effects on the transmitted flux stressing 
their importance for the reconstruction of the initial mass
density field. Viel {\it et al.}  (2002) implemented a variant 
of the BD97 model to simulate multiple systems of QSOs and found 
correlations in the transverse direction in agreement with
observations (see also Petry {\it et al.} 2002 for an analysis of 
the correlations in the transverse direction).

The BD97 model is based on the assumption that the low-column density
Ly$\alpha$ forest is produced by smooth fluctuations in the
intergalactic medium which arise as a result of gravitational growth of
perturbations. Since the fluctuations are only mildly non-linear,
density perturbations in the intergalactic medium $\delta_0^{\rm
IGM}({\bf x}, z)$ can be related to the underlying DM perturbations by
a convolution, which models the effects of gas pressure. In Fourier
space one has: \be \delta_0^{\rm IGM} ({\bf k}, z) = W_{\rm IGM}(k,z)
D_+(z) \delta_0^{\rm DM}({\bf k}) \ee where $D_+(z)$ is the growing
mode of density perturbations (normalised so that $D_+(0)=1$) and
$\delta_0^{\rm DM}({\bf k})$ is the Fourier transformed DM linear
over density at $z=0$. The low-pass filter $W_{\rm IGM}(k,z) = (1 +
k^2/k_J^2)^{-1}$ depends on the comoving Jeans length \be k_J^{-1}(z)
\equiv H_0^{-1} \left[ {2 \gamma k_B T_m(z) \over 3 \mu m_p \Omega_{0m}
(1 + z)}\right]^{1/2} \;,
\label{JL}
\ee where $k_B$ is Boltzmann's constant, $T_m$ the gas temperature,
$\mu$ the molecular weight and $\gamma$ the ratio of specific heats.
$\Omega_{0m}$ is the present-day matter density. The Jeans length
depends on density and temperature. Jeans smoothing is therefore an
adaptive smoothing of the density field. However, for simplicity 
in most practical implementations the comoving Jeans length is  
assumed to be constant  and computed for the mean density and the 
mean temperature $T_0$ at mean density $\delta=0$. As we will see
later this is a severe restriction which significantly affects the
results. 

Gnedin \& Hui (1998) adopt a different and more accurate expression
for the IGM filter $W_{\rm IGM}(k,z)$, which, however, does not allow
a simple matching with the non-linear regime (see also the discussion
in Section \ref{ZEL}). More accurate window-functions have also been
proposed by Nusser (2000) and Matarrese \& Mohayaee (2002). In what 
follows, we take $T_0(z) \propto 1+z$, which leads to a constant
comoving Jeans scale. This assumption should not be critical as the 
redshift intervals considered here are small.

Bi \& Davidsen adopt a simple lognormal (LN) transformation (Coles \&
Jones 1991) to obtain the IGM density in the mildly non-linear regime
from the linear density field, \be 1+\delta_{\rm IGM}({\bf x},z)=
\exp\left[\delta_0^{\rm IGM}({\bf x}, z) -
{\langle (\delta_0^{\rm IGM})^2 \rangle D_+^2(z) \over 2} \right]
\label{logn}
\ee
where $1+\delta_{IGM}({\bf x},z)=n_{\rm IGM}({\bf x},z)/{\overline
n}_{\rm IGM}(z)$ and  ${\overline
n}_{\rm IGM}(z) \approx 1.12 \times 10^{-5}
\Omega_{0b} h^2 (1+z)^3$ cm$^{-3}$.
The IGM peculiar velocity ${\bf v}^{\rm IGM}$ is related to the linear
IGM density contrast via the continuity equation. As in BD97, we 
assume that the peculiar velocity is still linear even on scales 
where the density contrast gets non-linear; this yields
\be
{\bf v}^{\rm IGM} ({\bf k},z) =  E_+(z) 
{i {\bf k} \over k^2} W_{\rm IGM}(k,z)\delta_0^{\rm DM}({\bf k})
\label{linvel}
\ee with $E_+(z) = H(z)f(\Omega_m,\Omega_\Lambda)D_+(z)/(1+z)$.  Here
$f(\Omega_m,\Omega_\Lambda) \equiv - d\ln D_+(z)/d \ln (1+z)$
(e.g. Lahav {\it et al.} 1991, for its explicit and general expression)
and $H(z)$ is the Hubble parameter at redshift $z$, \be H(z) = H_0
\sqrt{\Omega_{0m}(1+z)^3 + \Omega_{0{\cal R}}(1+z)^2 +
\Omega_{0\Lambda}} \ee where $\Omega_{0\Lambda}$ is the vacuum-energy
contribution to the cosmic density and $\Omega_{0{\cal
R}}=1-\Omega_{0m} - \Omega_{0\Lambda}$ ($\Omega_{0{\cal R}}=0$ for a
flat universe). The procedure to obtain 1D LOS random fields
from 3D random fields is described in BD97 and Viel {\it et al.}
(2002).  We will  explore below a more accurate mapping between the 
linear and non-linear gas density which uses  the PDF of the 
gas density obtained from hydro-dynamical simulations.

The neutral hydrogen density can be computed from the total density,
assuming photo-ionisation and taking the optically thin limit, $n_{\rm
HI}({\bf x},z) = f_{\rm HI}(T,\Gamma,Y) n_{\rm H}({\bf x},z)$, where
$n_{\rm H}=(1-Y)n_{\rm IGM}$ and $Y$ is the helium abundance by
mass. The photo-ionisation rate $\Gamma\equiv \Gamma_{12}\times
10^{-12}$s$^{-1}$ is related to the spectrum of ionising photons as
$\Gamma=\int_{\nu_{\rm th}}^\infty 4\pi J(\nu)/h_p\nu \sigma(\nu)
d\nu$, where $\sigma(\nu)$ is the photo-ionisation cross-section and
$h_P\nu_{\rm th}$ the hydrogen ionisation threshold ($h_P$ denotes
Planck's constant). $J(\nu) = J_{21} (\nu_0 /\nu)^{m} \times 10^{-21}
{\rm erg}~{\rm s}^{-1} {\rm Hz}^{-1} {\rm cm}^{-2} {\rm sr}^{-1}$ with
$\nu_{\rm th}$ the frequency of the HI ionisation threshold, and $m$ is
usually assumed to lie between 1.5 and 1.8. In the highly ionised case
($n_{\rm HI} \ll n_{\rm IGM}$) of interest here, one can approximate
the local density of neutral hydrogen as (e.g. Hui, Gnedin \& Zhang
1997)
\begin{eqnarray}
\frac{n_{\rm HI}({\bf x}, z)}{{\overline n}_{\rm IGM}(z)} 
&\approx& 10^{-5}\left({\Omega_{0b} h^2 \over 0.019}\right)
\left({\Gamma_{-12} \over 0.5}\right)^{-1} \left({1+z \over 4}\right)^3 \nonumber\\ 
& &\times\left(T({\bf x},z) \over 10^4 {\rm K} \right)^{-0.7} \left(1 + \delta_{\rm IGM}({\bf x},z) \right)^2.  
\label{ionization}
\end{eqnarray}

The temperature of the low-density IGM is determined by the balance
between adiabatic cooling and photo-heating by the UV background, which
establishes a local power-law relation between temperature and
density, $T({\bf x},z) = T_0(z) (1+\delta^{\rm IGM}({\bf
x},z))^{\gamma(z)-1}$, where both the temperature at mean density
$T_0$ and the adiabatic index $\gamma$ depend on the IGM ionisation
history (Meiksin 1994; Miralda-Escud\'e
\& Rees 1994; Hui \& Gnedin 1997; Schaye {\it et al.} 2000; Theuns {\it
et al.} 2001).

Given the neutral density, the optical depth in redshift-space at
velocity $u$ (in km s$^{-1}$) is \be \tau(u)= {\sigma_{0,\alpha}c\over
H(z)} \int_{-\infty}^{\infty}dy\, n_{\rm HI}(y) {\cal
V}\left[u-y-v_{\parallel}^{\rm IGM}(y),b(y)\right]
\label{tau} 
\ee where $\sigma_{0,\alpha} = 4.45 \times 10^{-18}$ cm$^2$ is the
hydrogen Ly$\alpha$ cross-section, $y$ is the real-space coordinate (in
km s$^{-1}$), ${\cal V}$ is the standard Voigt profile normalised in
real-space, and $b=(2k_BT/mc^2)^{1/2}$ is the thermal width. Velocity
$v$ and redshift $z$ are related through $d\lambda/\lambda=dv/c$, where
$\lambda=\lambda_0(1+z)$. For the low column-density systems considered
here, the Voigt profile is well approximated by a Gaussian: ${\cal
V}=(\sqrt{\pi} b)^{-1}\exp[-(u-y-v_{\parallel}^{\rm
IGM}(y))^2/b^2]$. As stressed by BD97 peculiar velocities affect the
optical depth in two different ways: the lines are shifted to a
slightly different location and their profiles are altered by velocity
gradients. In our modelling, we treat $\Gamma_{-12}$ as a free
parameter, which is varied the observed effective
opacity $\tau_{\rm eff}(z) = -\ln \langle \exp{(-\tau)}\rangle$ is
matched  (e.g. McDonald {\it et al.} 1999; Efstathiou {\it et al.} 2000) at the
median redshift of the considered range ($\tau_{\rm eff} = 0.12$ and
$\tau_{\rm eff}=0.27$ at $z=2.15$ and $z=3$, respectively, in our
case). We determine $\Gamma_{-12}$ by requiring that the ensemble
averaged effective optical depth is equal to the observed 
effective optical depth.  The transmitted flux is then 
simply ${\cal F}=\exp(-\tau)$.

\subsection{Improving the mapping from the linear 
dark matter density  to the non-linear gas density}
\label{improve}

In this subsection we describe how the mapping from the linear 
DM density to the non-linear gas density can be improved. 
As described in the previous section the semi-analytical  
modelling of Ly$\alpha$ forest spectra involves two main steps:
\begin{itemize}
\item smoothing of the linear DM density  
field to obtain a linear gas density field 
\item a local mapping from the linear to the non-linear gas 
density. 
\end{itemize}

\begin{figure*}
\resizebox{0.4\textwidth}{!}{\includegraphics{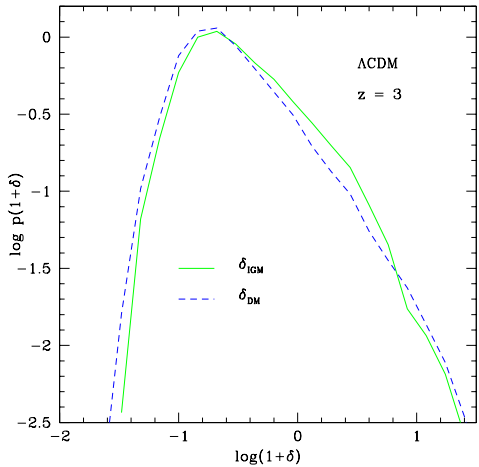}} 
\caption{Density distribution of the gas (IGM, continuous line) and
dark matter (DM, dashed line) obtained with SPH interpolation over 300
LOS extracted from hydro-simulations at $z=3$ for as $\Lambda$CDM
cosmology.}
\label{fig1}
\end{figure*}

\begin{figure*}
\resizebox{0.8\textwidth}{!}{\includegraphics{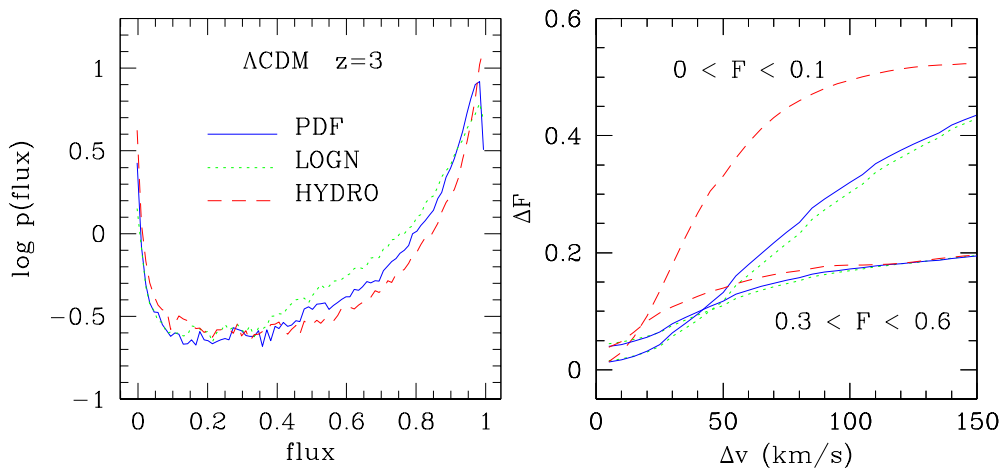}} 
\caption{Left panel: one-point function of the flux obtained from the
hydro-simulations (dashed line), lognormal model (dotted line) and
`improved' model (PDF, continuous line). A total number of $\sim 10^6$
pixels is used in the computation. All the spectra are computed at
$z=3$ for a $\Lambda$CDM model. Right panel: mean flux difference
$\Delta F(F_1,\delta v)$, for hydro-simulations (dashed line), lognormal
model (dotted line) and `improved' model (PDF, continuous line). Two
different flux intervals have been chosen which correspond to strong
absorbers ($0<F_1<0.1$) and to an intermediate strength lines
($0.3<F_2<0.6$).}
\label{flux}
\end{figure*}

\begin{figure*}
\resizebox{0.8\textwidth}{!}{\includegraphics{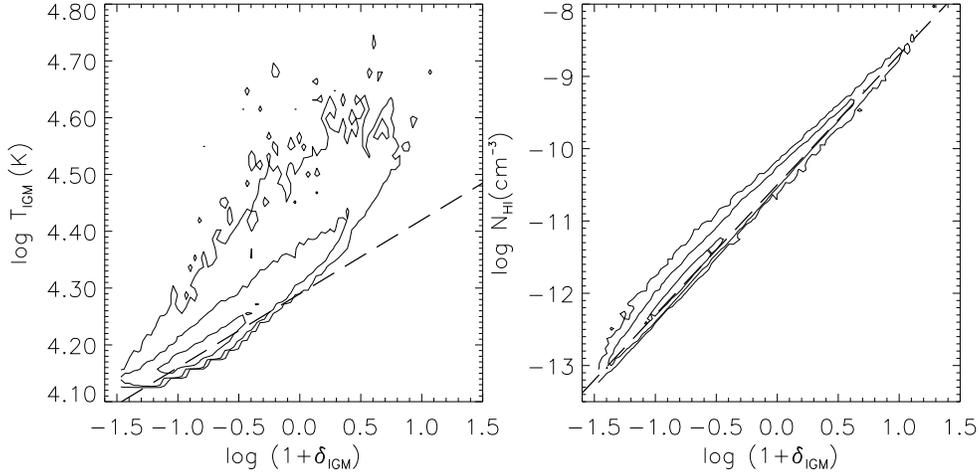}} 
\caption{Left panel: contour plot of $\log(T_{IGM})$
vs. $\log(1+\delta_{IGM})$, at $z=3$. The dashed line is obtained by
setting $\gamma=1.2$. Right panel: contour plot of $\log(N_{HI})$
(comoving neutral hydrogen density in cm$^{-3}$)
vs. $\log(1+\delta_{IGM})$ obtained from the $z=3$ output of the
hydro-simulation. The dashed line is given by
eq. (\ref{ionization}). Levels of contours increase by a factor 10.  (a
total number of $3\times 10^5$ points is used in the computation).}
\label{fit1}
\end{figure*}

In this section we will substitute the log-normal mapping 
used by BD97 for the second step by a rank-ordered mapping 
from the linear gas density field to the PDF of the gas 
density  in the hydro-simulation.  Figure \ref{fig1} compares 
the PDF of the DM 
density field obtained using SPH interpolation and that of the gas 
density field. Pressure forces push the gas out into the surrounding 
voids (\lq Jeans smoothing\rq) and for this reason the PDF of the gas 
drops below that of the DM density at low densities. At high density
the  gas is converted into stars and the PDF of the gas density drops
again  below the PDF of the DM density. As a result the  PDF of the 
gas density is more peaked  (Theuns, Schaye \& Haehnelt 2000).
Figure \ref{flux} compares the resulting probability distribution of
the flux for  simulated spectra using the lognormal model our  
`improved' model  (hereafter PDF model) and the numerical simulations. 
For this comparison we have produced simulated spectra  for the 
lognormal model and the PDF model   with the same cosmological parameters as 
in our hydrodynamical simulation. We further  
imposed a power law temperature-density relation with
$T_0=10^{4.3}$ K and $\gamma=1.2$, which fits the simulation well
(Figure \ref{fit1}).  The simulated spectra have larger length ($\sim 16000
\vel$) than the spectra extracted from hydro-simulations ($\sim 1400
\vel$), but are obtained with the same spectral resolution,  approximately
of $2 \vel$.  We take into account that the hydro-simulations is 
missing large scale power by  applying a
similar cut-off in the power spectrum used to calculate the
semi-analytical spectra. This should allow a fair comparison. 
The peculiar velocity field is again assumed
to be given by linear theory (Eq.~\ref{linvel}).  All spectra are
scaled to $\tau_{eff}=0.27$ at $z=3$ and no noise is added.

The probability distribution of the flux for our PDF model 
agrees significantly better with that of the spectra obtained 
from  our hydro-dynamical  simulation.  There are still 
small but significant differences. While the agreement for this 
1 point statistic seems reasonable the agreement becomes very poor 
for some 2-point statistics of the flux.  The function $P(F_1,F_2,\Delta
v)dF_1\,dF_2$ is the probability that two pixels separated in the
spectrum by a velocity difference $\Delta v$, have transmitted fluxes
in intervals $dF_1$, $dF_2$ around $F_1$ and $F_2$, respectively. In
the right panel of Figure \ref{flux} we show  the mean flux
difference, 
\begin{equation}
\Delta F(F_1,\delta v)=\int P(F_1,F_2,\delta v) (F_1-F_2)
dF_2\,,
\label{eq:deltaF}
\end{equation}
for 2 values of $F_1$, as a function of $\Delta v$ (see
Miralda-Escud\'e {\it et al.}  1997, Theuns {\it et al.}  2000, Kim
{\it et al.} 2001).
The results for the lognormal model and for the PDF model are almost
identical, which is not surprising since we checked that the two functions that
map the linear density field into  the non-linear density field are
similar.

For strong absorbers ($0\le F\le 0.1$) the agreement between the
numerical simulation and our models which rely on an analytical
description of the DM density field is very  poor. {\it What could be the
reason for this discrepancy?}
We have made the following major  simplifications: 
{\it (i)} the peculiar velocity field is assumed to be linear; {\it (ii)}
we do not simulate the intrinsic scatter present in the relation
$T_{IGM}$ vs. $\delta_{IGM}$; {\it (iii)} we recover directly the
neutral hydrogen density along the LOS by assuming
eq. (\ref{ionization}), which is valid only under some assumptions
(see Subsection \ref{sub1}); {\it (iv)} we input only the 1-point
probability distribution function of the IGM, without taking into
account the higher order moments of the distribution; {\it (v)} the
correlations are assumed to be those predicted by linear theory,
modified by the non-linear mapping, while the correlations of the
hydro-simulation are different.

Concerning point $(i)$ it has been shown by Hui {\it et al.}  (1997)
that the peculiar velocity field can affect the shape of the
absorption features, while it has very small effect on the column
density distribution functions of the lines. We have run the
calculated spectra varying  the peculiar velocity field and
have not found a strong dependence of the PDF of the flux on 
the peculiar velocity quantity. This means that this statistics is 
mainly influenced by the underlying density field. We have compared the 
PDF of the peculiar velocity field predicted by linear theory with 
the peculiar velocity found in the $z=3$ output of 
hydro-simulation for the IGM and we have found that the 
differences are not big. The assumption of a linear velocity field 
should therefore be a good approximation.

If we try to simulate the intrinsic scatter in the relation
$T-\delta$, left panel of Figure \ref{fit1}, for example by using as
input a higher order polynomial fit instead of the power-law relation,
the improvement in the PDF of the flux is negligible.  Provided we
set the same $T_0$ of the simulations and a `reasonable' value of
$\gamma$, the power-law equation of state is a good approximation and
the scatter can be neglected.

The same considerations are valid for point $(iii)$. In fact, we have
found that eq. (\ref{ionization}) is in good agreement with
hydro-simulations (Figure \ref{fit1}, right panel), showing that the
approximations we made (optical thin limit, high ionisation state) are
not affecting the results significantly.

We thus believe that point $(iv)$ and $(v)$ are the main simplification
involved in the model. This is somewhat expected as in the definition
of the optical depth, eq. (\ref{tau}), the summation is done over
nearby pixels, so it is influenced by the spatial correlations, which
are assumed to be those predicted by the linear theory, modified by the
local non-linear mapping.  
In the following Sections we will thus turn to  models that are based on the DM density
field obtained directly from a $N$-body simulation (Fig.~\ref{2PDFall}
below). These models reproduce the flux correlation for strong
absorbers much better. This demonstrates that correlations in the
density field  introduced by the non-linear evolution 
are  indeed responsible for the shape of strong lines. These are not not 
reproduced accurately by our local mapping from the linear to
non-linear gas density.  The marked differences for strong 
absorbers are not unexpected as the $z=3$ output of hydro-simulation 
contains very strong absorption systems from fully collapsed 
objects which are  not related to the  Ly$\alpha$ forest. 
The spectra of the LOGN model have been produced using
the Jeans length at the mean temperature and with a fixed value of
$\gamma$. If we use other values for the temperature $T_0$ and for
$\gamma$ we can obtain a better agreement in other flux intervals. In
Figure \ref{flux} the range of flux values in which the agreement with
the flux PDF of the hydro-dynamical simulation and in $\Delta F$ is
good corresponds to intermediate absorbers ($0.2<F<0.4$), this is 
in part determined by the requirement of having a fixed 
$\tau_{eff}=0.27$ for the ensemble of simulated spectra in each model, 
which scales the neutral hydrogen fraction.

\section{Models using Numerical simulations of the dark matter distribution}
\label{allmeth}
\subsection{Zel'dovich modelling of the gas distribution}
\label{ZEL}
In this Section we present  models of the gas distribution based on 
a modified filtering of the initial conditions of the dark matter density
field. The first method is based on the truncated Zel'dovich
approximation (TZA). Among the possible approximations, TZA has
been found to produce the best agreement with N-body results (see Coles
{\it et al.}  1993, Melott {\it et al.} 1994, Sathyaprakash {\it et
al.} 1995).  Hui {\it et al.} (1997, hereafter HGZ) showed that the
TZA, with an appropriate smoothing and with a recipe which allows to
convert density peaks into absorption lines, successfully reproduces
the observed column density distribution of the Ly$\alpha$ forest lines
over a wide range of $N_{HI}$. Gnedin \& Hui (1998) presented a more
accurate semi-analytical model of the Ly$\alpha$ forest by combining a
particle mesh solver modified to compute also an effective potential
due to gas pressure. They showed that a particle mesh solver, with an
appropriate filtering of the initial condition, can be used to model
the low density IGM.  All these different methods have the advantage of
being much faster than a hydro-simulation.  

We are now going to make a comparison, LOS by LOS,
between the simulated gas distribution and the effective gas
distribution of the $z=3$ output of the hydro-simulation. Our
simulation is significantly larger and with higher resolution than those
used in previous studies.

Following Hui et al. (1997), we define a non-linear wavenumber $k_{nl}$
(see also Melott {\it et al.} 1994): \be
D^2_+(z)\int_0^{k_{nl}} P(k)d^3{\bf{k}}=1,
\label{variance}
\ee where $D_+(z)$ is the linear growth factor for the density
perturbations and $P(k)$ the linear power-spectrum.  We filter the
linear density field with a Gaussian window
$W(k,k_s)=\exp(-k^2/2k_s^2)$, with $k_s\sim 1.5 k_{nl}$.  At this point
we displace particles from the initial Lagrangian coordinates $\bf{q}$
according to the truncated Zel'dovich approximation: \be
{\bf{x}}({\bf{q}},z) = {\bf{q}} +
D_+(z)\nabla_{\bf{q}}\phi_f({\bf{q}}), \ee with $\phi_f(q)$ the initial
{\it filtered} velocity potential (see \eg Coles {\it et al.} 1995),
with $\phi(\bf{k})$ the actual potential used in the initial conditions
of the simulation. The filtering is intended to prevent shell-crossing.
To mimic baryonic pressure, we smooth the initial density field with a
Gaussian window $\exp(-k^2/2k_J^2)$, with $k_J$ given by
eq. (\ref{JL}), as in HGZ. The final filtering scale is effectively the
smaller of $k_s$ and $k_J$. In our case, $k_s \sim 5.5 \mpc^{-1}$ and
$k_J\sim 7\mpc^{-1}$, these wave numbers correspond to scales of
$\lambda_s\sim 1.14\, \mpc$ and $\lambda_J\sim 0.9\, \mpc$,
respectively.  This means that the amount of filtering, before
displacing particles, is given by the condition of
eq. (\ref{variance}).

We also outline here a further approach based on the assumption that the
displacement between DM and IGM particles is `Zel'dovich-like'. We
refer to this method as the Zel'dovich Displacement (ZD method), while
the method in which we assume a perfect tracing between the DM and IGM
density field will be referred as the DM method.
The difference, at redshift $z$, between the Eulerian position of a gas
and a DM particle which have the same Lagrangian coordinate $\bf{q}$,
according to the Zel'dovich approximation, is:
\be
\Delta{\bf{x}}({\bf{q}},z)=D_+(z)[\nabla_{\bf{q}}\psi_{IGM}({\bf{q}},z)
-\nabla_{\bf{q}}\phi_{DM}({\bf{q}})],
\label{disp}
\ee where $\psi_{IGM}$ and $\phi_{DM}$ are the IGM and DM velocity
potentials. In Fourier space we have (Matarrese \& Mohayaee 2002)
$\psi_{IGM} ({\bf{k}},z)=W_{IGM}({\bf{k}},z)\phi_{DM}({\bf{k}})$ and
from eq. (\ref{disp}) we get: \be
\Delta{\bf{x}}({\bf{k}},z)=D_+(z)[W_{IGM}-1]i
{\bf{k}}\phi_{DM}({\bf{k}}).  \ee The above equation shows that for $k
\ll k_J$ $W_{IGM}\to 1$, $\Delta{\bf{x}}({\bf{k}},z)\to 0$, while for
$k \gg k_J$, $W_{IGM}\to 0$, $\Delta{\bf{x}}({\bf{k}},z)\to - i
{\bf{k}}\phi_{DM}(\bf{k})$. In the first limit, Jeans smoothing is not
effective and the IGM traces the DM, in the second limit the effect of
gas pressure prevents large displacements of baryons from their initial
location.  According to this method, the displacement between DM and
IGM particles with the same Lagrangian coordinate is given by a
filtering of the initial DM density field.  We calculate this
displacement at $z=3$ with the filter proposed by Gnedin \& Hui (1998),
\ie instead of $W_{IGM}$ we use $W_{ZD}=\exp(-k^2/2k_{ZD}^2)$, with
$k_{ZD}\sim 2.2 k_J$. This choice is motivated by the fact that a
Gaussian filter gives an excellent fit to baryon fluctuations for a
wide range of wave numbers. The filtering scale $k_{ZD}$ is the value
predicted by linear theory at $z=3$, assuming that reionization takes
places at $z \sim 7$ (Gnedin \& Hui 1998). The location of the gas
particle is then found be adding this displacement to the {\it actual
position} at $z=3$ of the DM particle with the same Lagrangian
coordinate.

\begin{figure*}
\resizebox{1.0\textwidth}{!}{\includegraphics{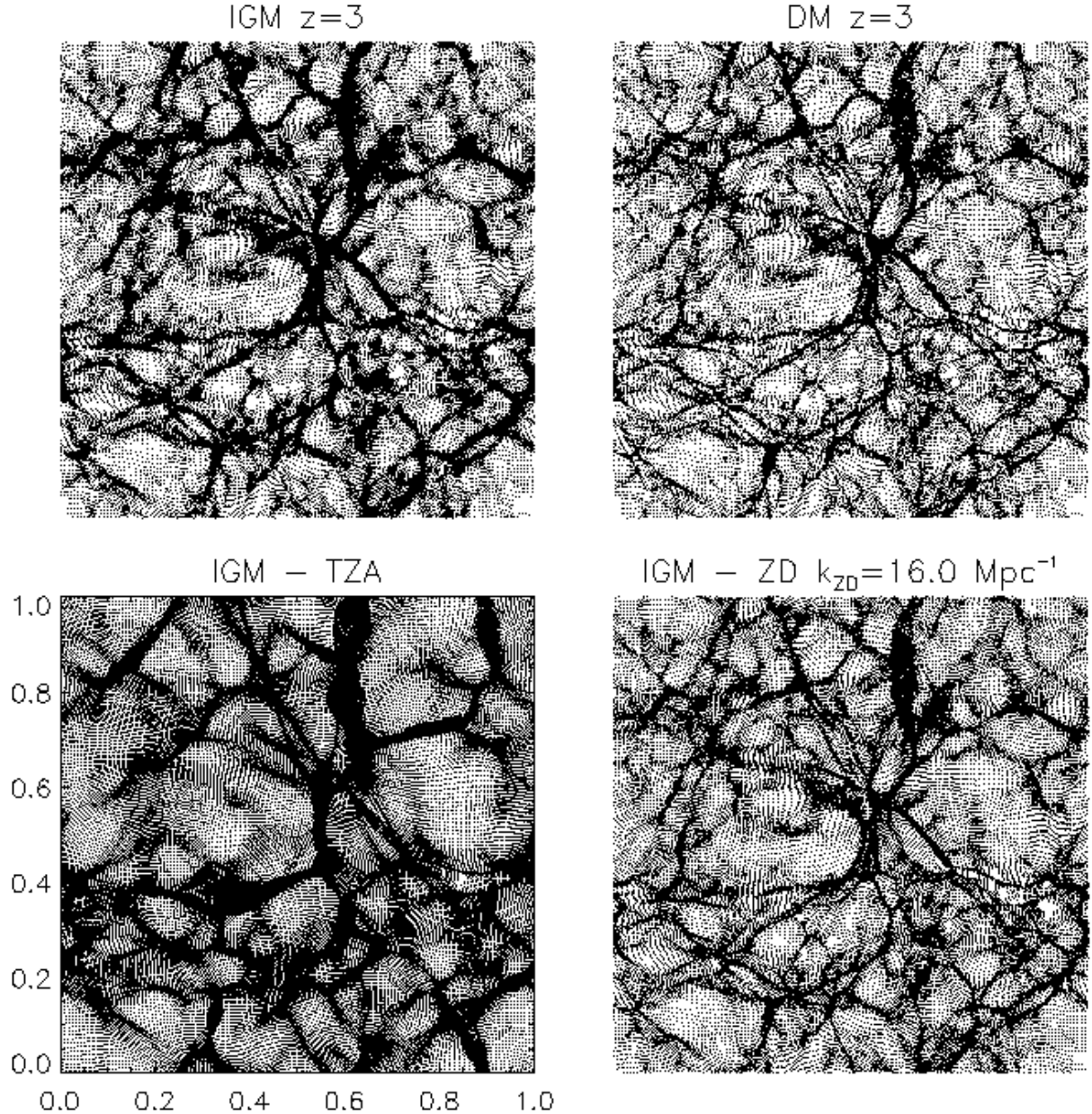}} 
\caption{Slices along the z-axis of thickness $\sim 0.1 h^{-1} \,{\rm comoving} ~\mpc$ (coordinates in normalised units). Top panels: IGM distribution (left)
and DM distribution (right), from the $z=3$ output of the $\Lambda$CDM
model. Bottom panels: Zel'dovich modelling of the IGM distribution
from the initial conditions of hydro-simulation (left panel, TZA),
Zel'dovich displacement added to DM particles to mimic baryonic
pressure (right panel, ZD).}
\label{slice}
\end{figure*}

In Figure \ref{slice} we show four panels, which are slices of
thickness $\sim 0.1 ~h^{-1}$ comoving Mpc at the same position along
the z-axis, for the IGM and the DM distribution of the hydro-simulations
(top panels), the TZA field (bottom left panel) and the IGM field
obtained with ZD method (bottom right).

By comparing the top panels, one can see that the gas is more diffuse
than the dark matter. TZA reproduces the main filaments but we know
that the agreement will be better in the low-density regions.  ZD
(with a filtering at $k_{ZD}\sim 16
\mpc^{-1}$) seems promising, at least by eye. This method allows some
diffusion around the dark matter to reproduce the gas distribution.  A
more quantitative comparison is needed to see how good our IGM density
field is compared to the hydro one. We compare LOS by LOS the SPH
interpolated IGM density fields with the `true' IGM field of the
hydro-simulation. This test is pretty severe as we are not going to
filter the density fields and the comparison is made `pixel-by-pixel'.

\begin{figure*}
\resizebox{0.8\textwidth}{!}{\includegraphics{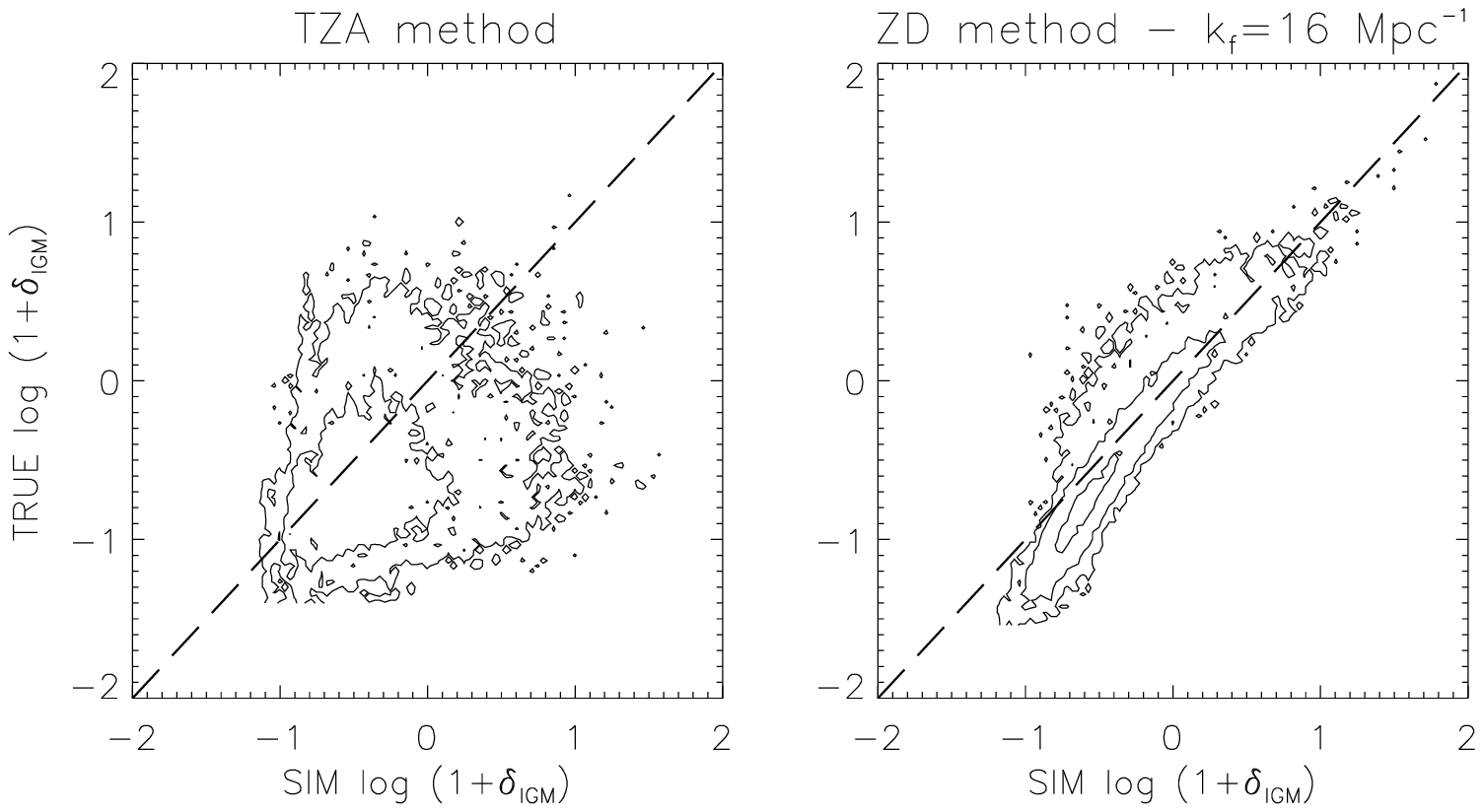}} 
\caption{Contour plots of the true IGM over density derived from the hydro-simulation 
and the simulated IGM over densities obtained with the truncated
Zel'dovich approximation (TZA, left panel) and the `Zel'dovich
displaced' method (ZD, right panel), at $z=3$. $k_f$ for ZD is $\sim
16 \mpc^{-1}$, $k_s$ for TZA is $\sim 5.5 \mpc^{-1}$. In both the
panels, the number density of the elements increases by an order of
magnitude with each contour level (a total number of $3\times 10^5$
points are used in the computation). }
\label{compare}
\end{figure*}

\begin{figure*}
\resizebox{0.4\textwidth}{!}{\includegraphics{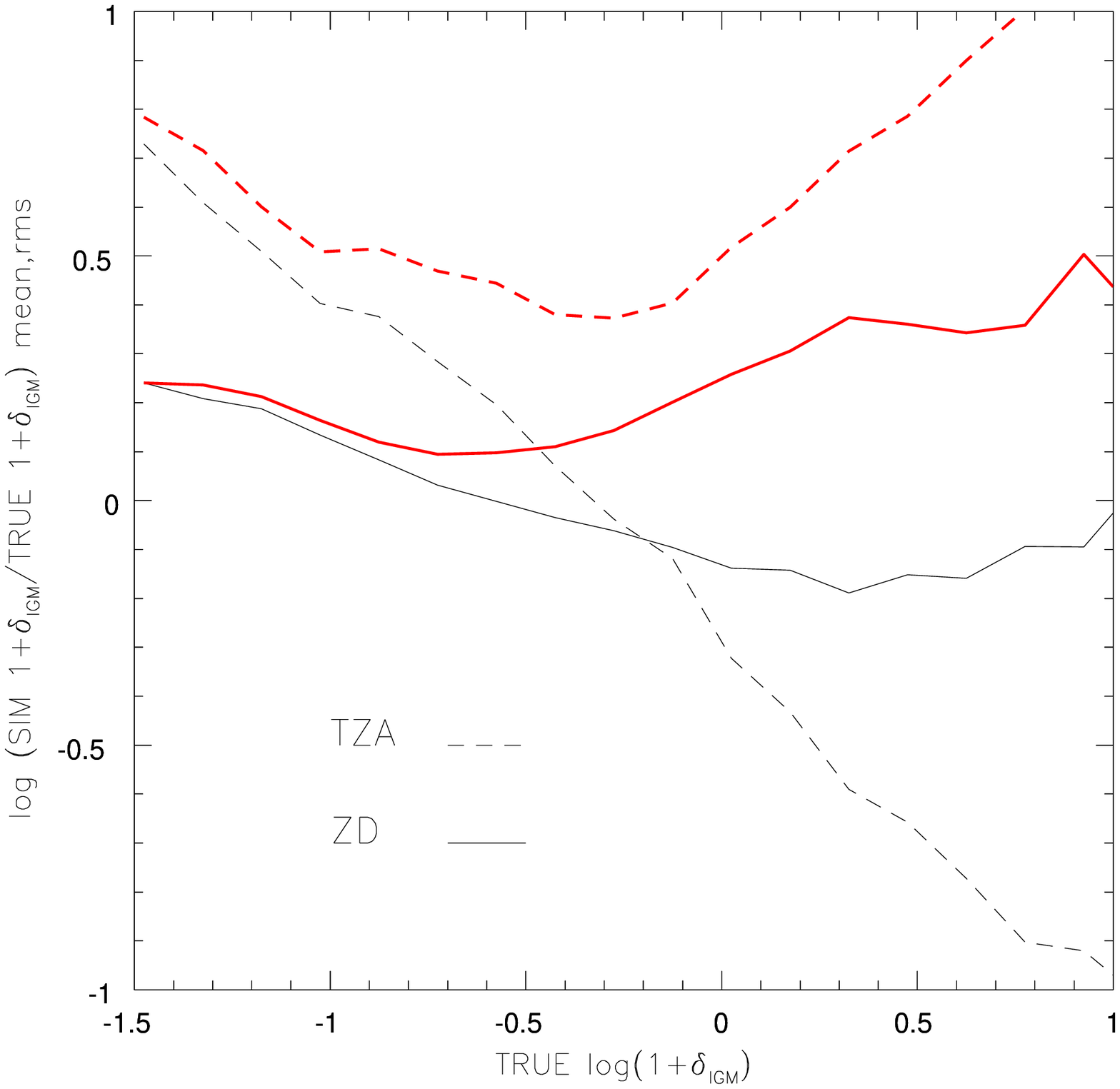}} 
\caption{Average and rms value of the difference between 
$\log {\rm SIM} (1+\delta_{IGM})$ and $\log {\rm TRUE}
(1+\delta_{IGM})$, with the TZA (dashed) and ZD method
(continuous). Thick lines represent the rms values, thin lines the average.}
\label{scatterHG}
\end{figure*}

The results are shown in Figure \ref{compare} where we compare the
amount of scatter predicted by the TZA approximation (left panel) and
the ZD method (right panel). As expected, the test without any
filtering on the {\em final} density field shows that the TZA tracks
the simulation, but the scatter is very large. The ZD method agrees
better with the simulation, but the scatter is still rather
significant. A more detailed comparison is shown in Figure \ref{scatterHG},
where we plot the mean, and scatter around the mean, as a function of density.

Clearly ZD works significantly better than TZA, the average value is
better reproduced and also the scatter is significantly smaller. 
The inadequacy of the TZA was also shown by Bond \& Wadsley (1997).
The amount of scatter and the average value found with the ZD method are
consistent with the results of Gnedin \& Hui (1998; see their Figure
4).

\subsection{Using the mean  $\delta_{DM}-\delta_{IGM}$ relation
of the hydro simulation to predict the gas distribution from the DM 
distribution.}
\label{imple}

The results of the previous section have demonstrated 
that `filtering' techniques on the initial conditions 
do not result in density fields that agree well with
hydro-simulations, at least for a `pixel-by-pixel'
comparison with a high-resolution simulation.

In this section we present an alternative approach that 
starts from the actual DM density distribution of the numerical 
simulation and ``predicts'' the gas distribution using 
a fit to the mean relation between gas and DM density. 
In this way the displacement of the gas with respect to the 
dark matter is modelled statistically in {\it real space}. 
The techniques of the previous Section are all based on 
different filtering schemes which smooth the IGM density 
field over a constant scale set by the  Jeans length {\it at
the mean density}. This filtering is done in {\it Fourier space} and
leads to  equal smoothing of all  dark matter structures
independent of their density. As discussed above this strong 
simplification is responsible for the rather strong discrepancies
with the gas distribution in our hydro-dynamical simulation.

We compute IGM and DM over densities, IGM peculiar velocity and IGM
temperature obtained with SPH interpolation (see Appendix \ref{SPH} for
details). From the computed IGM and DM density profiles along different
LOS we argue that a simple relation between the DM and the gas IGM
density fields does not exist (see also Cen {\it et al.} 1994, Croft {\it et al.}
1998; Zhang {\it et al.} 1998). For a wide range of moderate over
densities, \ie $- 0.75 \mincir \delta \mincir 5$, the PDF of the gas
distribution lies above that of the DM (Fig.~\ref{fig1}): this of
course does not determine the relation between gas and dark matter, but
shows that we should expect a significant number of regions in which
$\delta_{IGM}$ is larger than $\delta_{DM}$.

This is what we find in the hydro-simulations in which there are
regions at high $\delta$ where the gas is more concentrated than the
dark matter. At low $\delta$ there is an almost perfect agreement
between the gas and the dark matter distribution, while at moderate and
large over densities the relation between gas and dark matter is not
unique. The highest density peaks are usually related to peaks in the
temperature of the gas of and strong gradients in its peculiar
velocity, suggesting the presence of a shock. However, it can happen
that a region shows a gas density more peaked than the DM, while the
velocity field and the temperature show no particular behaviour. This
means that modelling of the gas distribution using only the information
along the LOS will not be very accurate on a point-to-point basis, but
we might nevertheless hope to obtain a model for the gas distribution
which has the correct properties in a statistical sense.

\begin{figure*}
\resizebox{0.6\textwidth}{!}{\includegraphics{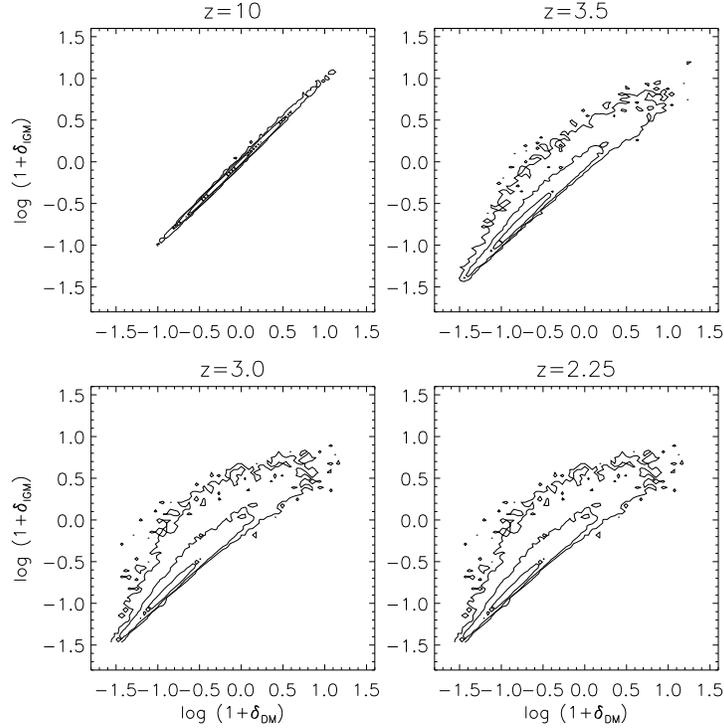}} 
\caption{Scatter plots of $\log(1+\delta_{IGM})$ vs. $\log(1+\delta_{DM})$, 
at $z=10$ (top left), $z=3.5$ (top right), $z=3$ (bottom left) and
$z=2.25$ (bottom right) of the hydro-simulation. 30 LOS of $2^{10}$ pixels
are reported here. Levels of contours increase by a factor of 10.}
\label{allscatter}
\end{figure*}

We start by plotting the values of $\delta_{IGM}$ vs. $\delta_{DM}$
obtained from the simulation at different redshifts. Figure
\ref{allscatter} demonstrates that the scatter between the IGM and DM
densities increases with decreasing redshift, as expected. At $z=10$,
there is almost perfect agreement  between the two fields, but when
cosmic structures get non-linear the physics becomes more complex. In
addition, reionization occurs in our simulation at $z\sim 6$, so before
that the simulations do not resolve the then much smaller Jeans length. At
later redshifts, the effect of the baryonic pressure can be seen. For
under dense regions, the IGM density tends to be higher than the DM
density, because pressure is pushing gas into the low density voids.
For larger values of $\delta_{DM}$ the scatter increases and gas can be
either denser or less concentrated than the dark matter, depending on
star formation which sets-in in the simulations at over densities $\ge
80$ (see Aguirre, Schaye and Theuns 2002 for a more detailed
description of the star formation recipe adopted).

\begin{figure*}
\resizebox{0.5\textwidth}{!}{\includegraphics{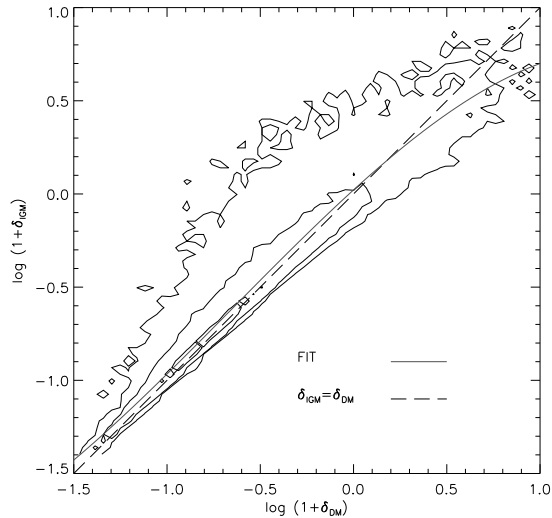}} 
\caption{IGM density contrast vs. DM density contrast, 
at $z=3$ for the $\Lambda$CDM model. The continuous line is the 3rd order
polynomial fit, the dashed line is obtained by setting
$\delta_{IGM}=\delta_{DM}$.}
\label{fitscatter}
\end{figure*}

We fit the relation between gas and DM density with a 3rd order
polynomial; the results are shown in Figure \ref{fitscatter}, where the
range of $\delta_{DM}$ plotted is the one of interest for the
Ly$\alpha$ forest, $-1\mincir \delta \mincir 6$. If we set
$y=\log(1+\delta_{IGM})$ and $x=\log(1+\delta_{DM})$ the fitting
function is:
\begin{eqnarray}
y &=&
0.02^{\pm 0.03}+0.91^{\pm 0.05}\,x-0.15^{\pm 0.05}\,x^2
\nonumber \\ &-& 0.08^{\pm 0.04}\,x^3, \,\,\,\,\,\,\,\,\, -1.5\le
x \le 1.0, \
\label{fitequation}
\end{eqnarray}
where the errors reported are the $1\sigma$ uncertainties of each
coefficient.  The same function has been found to be a reasonable good
fit at $z=4$ and $z=2$ as well, so it can be used to simulate the gas
distribution in this redshift range for a $\Lambda$CDM model. From
Figure \ref{fitscatter} one can see that the fitting function predicts
that the gas is on average more concentrated than the dark matter for a
wide range of $\delta_{DM}$, although significant scatter is
present.

\begin{figure*}
\resizebox{0.8\textwidth}{!}{\includegraphics{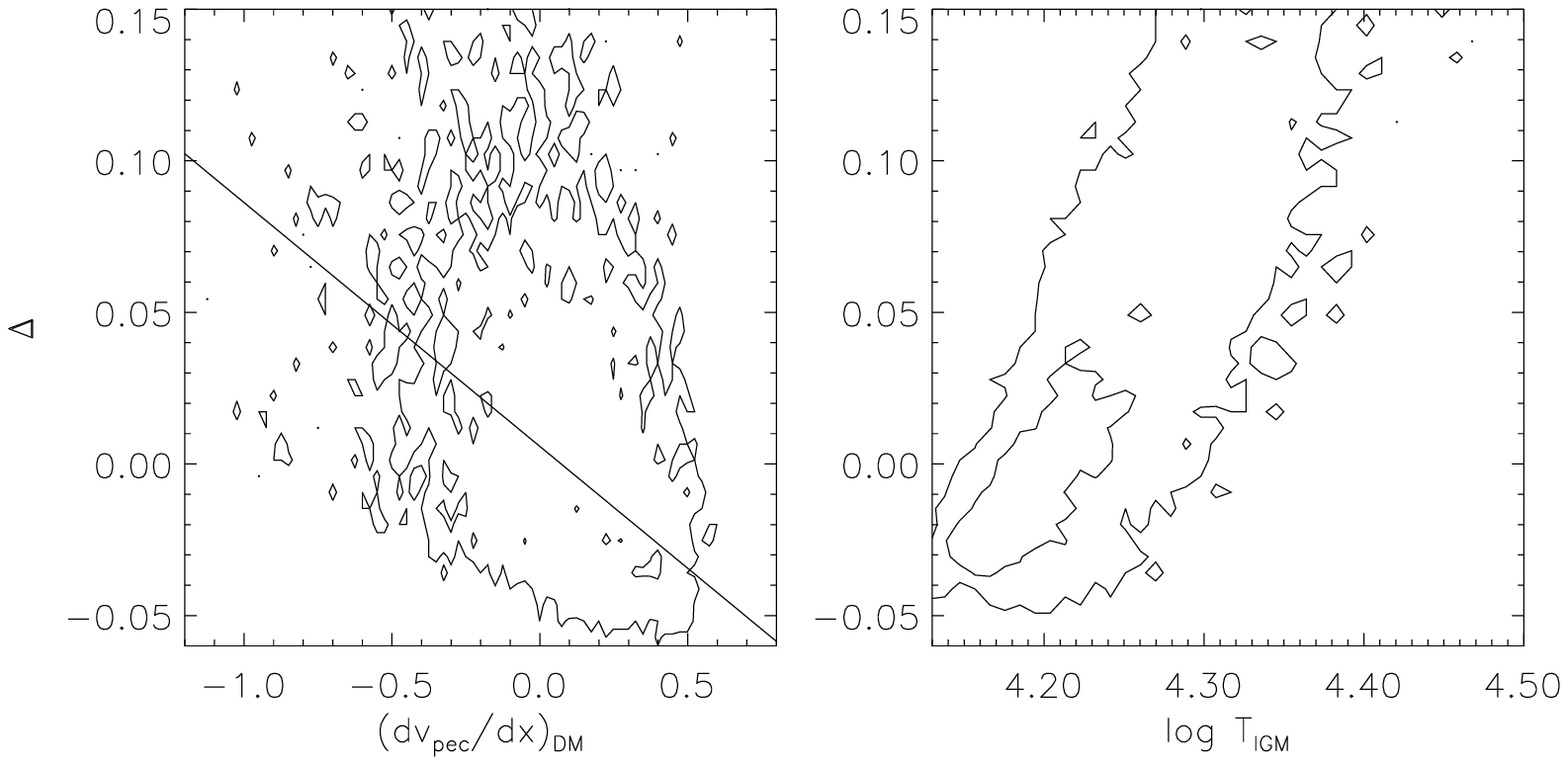}} 
\caption{Scatter plots of the dark matter peculiar velocity gradient 
along the LOS vs. $\Delta$ (left panel) and temperature of the IGM
vs. $\Delta$ (right panel), where $\Delta=\log(1+\delta_{IGM})-{\rm fit}$,
i.e. the difference between the true value and the value obtained
with the fit, (see eq. \ref{fitequation}). Continuous line in the left
panel represents the fit we have used in simulating the IGM density field.
Weak correlations are present. Levels of contours increase by a factor
of 10.}
\label{scatt_analysis}
\end{figure*}
{\it Is it possible to reduce this scatter using other
information?}

In Figure \ref{scatt_analysis} (left panel) we plot  the
peculiar velocity gradient of the dark matter density fields vs. the
difference $\Delta$ between the `true' value of $\log(1+\delta_{IGM})$
and the `fit' value obtained with eq. (\ref{fitequation}). Both these
quantities have been smoothed over a scale of $\sim 100 \vel$ to reduce
noise. $\Delta$ is weakly anti-correlated with the dark matter peculiar
velocity gradient, with regions with a negative gradient lying above
the mean fit and hence have $\Delta >0$. On average, a negative
velocity gradient indicates that the gas is being compressed and so 
may be undergoing moderate or strong shocks. As a consequence, the gas 
is also being heated, and this introduces the correlation between $\Delta$ and
the temperature of the gas (Fig.~\ref{scatt_analysis}, right
panel). This correlation is stronger, because the gas temperature is a
more direct indicator of a shock. Conversely, negative $\Delta$, \ie
points below the fitting function, are related to colder regions of gas
and small positive gradients of the dark matter peculiar velocity
field. As expected, the bulk of the points is in quiet regions with
$dv_{pec}/dx\sim 0$ and temperatures between $10^{4.2}$ K and
$10^{4.3}$ K.

These results suggest that we can indeed model the gas distribution
that corresponds to a given dark matter density and peculiar velocity
distribution statistically.  We use the fit of eq. (\ref{fitequation})
to predict the  gas density field from the dark matter density field. 
The modelling can be further improved by taking the peculiar velocity 
of the dark matter into account. For this purpose we fit the 
relation shown in the left panel of Figure \ref{scatt_analysis}
with a 2nd order polynomial because the correlation is weak. 
Models obtained using the  density-density fit and models using 
a two parameter density-density   plus the density-peculiar
velocity fit will be referred to as FIT-1 or FIT-2, respectively.

\begin{figure*}
\resizebox{0.4\textwidth}{!}{\includegraphics{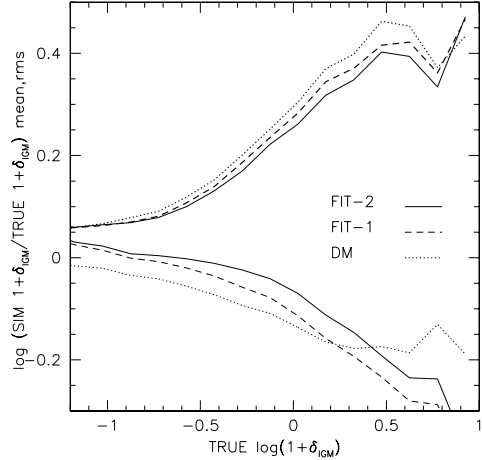}} 
\caption{Average (thin lines) and rms (thick lines) value of the
difference between $\log {\rm SIM} (1+\delta_{IGM})$ and $\log {\rm
TRUE} (1+\delta_{IGM})$, for the FIT-2 (continuous lines), FIT-1
(dashed lines) and DM (dotted lines) methods.}
\label{compareall}
\end{figure*}

\begin{figure*}
\resizebox{0.8\textwidth}{!}{\includegraphics{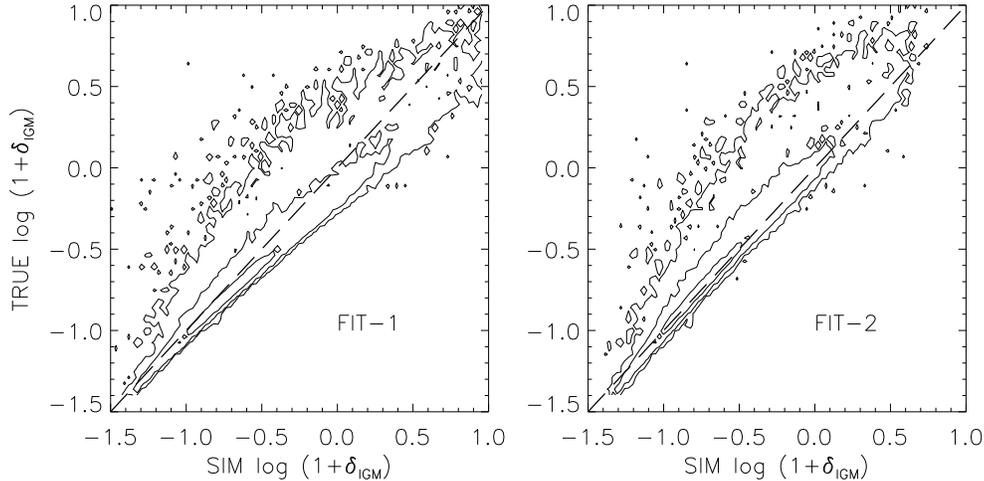}} 
\caption{Left panel: true versus simulated (FIT-1 model, 
Eq.~\ref{fitequation}) $\delta_{IGM}$ at $z=3$ for the $\Lambda$CDM
simulation plotted vs simulated $\delta_{IGM}$ (FIT-1 model) obtained.
with the fit of eq. (\ref{fitequation}). Right panel: true
$\delta_{IGM}$ versus that obtained using FIT-2 model, improved with
the fitting of the scatter (see Figure \ref{scatt_analysis}). Contour
levels are off-set by 1 dex, and are based on $3\times 10^5$
points.}
\label{scatter_los1}
\end{figure*}

The fitting function of eq. (\ref{fitequation}) gives a good
approximation to the gas distribution for a $\Lambda$CDM model in the
redshift range $2\mincir z \mincir 4$. The  same technique can be
applied to other cosmological models and different redshift ranges. We
will now  make a LOS by LOS comparison  for the the gas density
distribution of the two models FIT-1 and FIT-2 to that of our 
hydro-dynamical simulation. In Figure \ref{scatter_los1} we
show the scatter plots for the gas (over)densities for the two models. 
Figure \ref{compareall} shows the scatter in terms of the
standard and mean deviation for both models. The simple  DM model where
the gas is assumed to trace the dark matter exactly is also shown 
(thick, thin and dotted lines, respectively). 

Figure \ref{scatter_los1} illustrates the improvement of using the
fitting procedure of model FIT-2. The $\chi^2$ for these fits, where
$\chi^2\equiv N_{bin}^{-1}\sum_{i=1}^{N_{bin}} \frac{(\delta_i -
\delta)^2}{\sigma_i^2}$ ($N_{bin}=25$ is the number of bins) are 0.24,
0.21 and 0.19 for the FIT-1, FIT-2 and DM, respectively, showing that
the scatter is effectively reduced with the second method but still in
the total interval $-1.5<\log(1+\delta_{IGM})<1$.

When the density range is constrained to be in the interval $-1
<\log(1+\delta_{IGM})<0.6$, relevant for the Lyman$\alpha$ forest, we
obtain $\chi^2$ of 0.06, 0.03 and 0.10 for the FIT-1, FIT-2 and DM
model respectively, illustrating the improvement of the fitting
procedure. Comparing with Figure \ref{scatterHG}, we see that the FIT-2
method is significantly better than ZD in reproducing the mean values
for $\log(1+\delta_{IGM})\mincir 0.5$ (however, for larger over densities
ZD shows a better agreement than FIT-2 or FIT-1),
while the rms values are basically equivalent. If we apply the fitting
procedure in the range $\delta_{IGM} \mincir 1$ which represents
the bulk of the IGM, the mean and rms deviation between the fitted and
true IGM densities are smaller than 10 and 30 per cent, respectively.

\subsection{Jeans smoothing of the evolved DM density field} 
\begin{figure*}
\resizebox{0.8\textwidth}{!}{\includegraphics{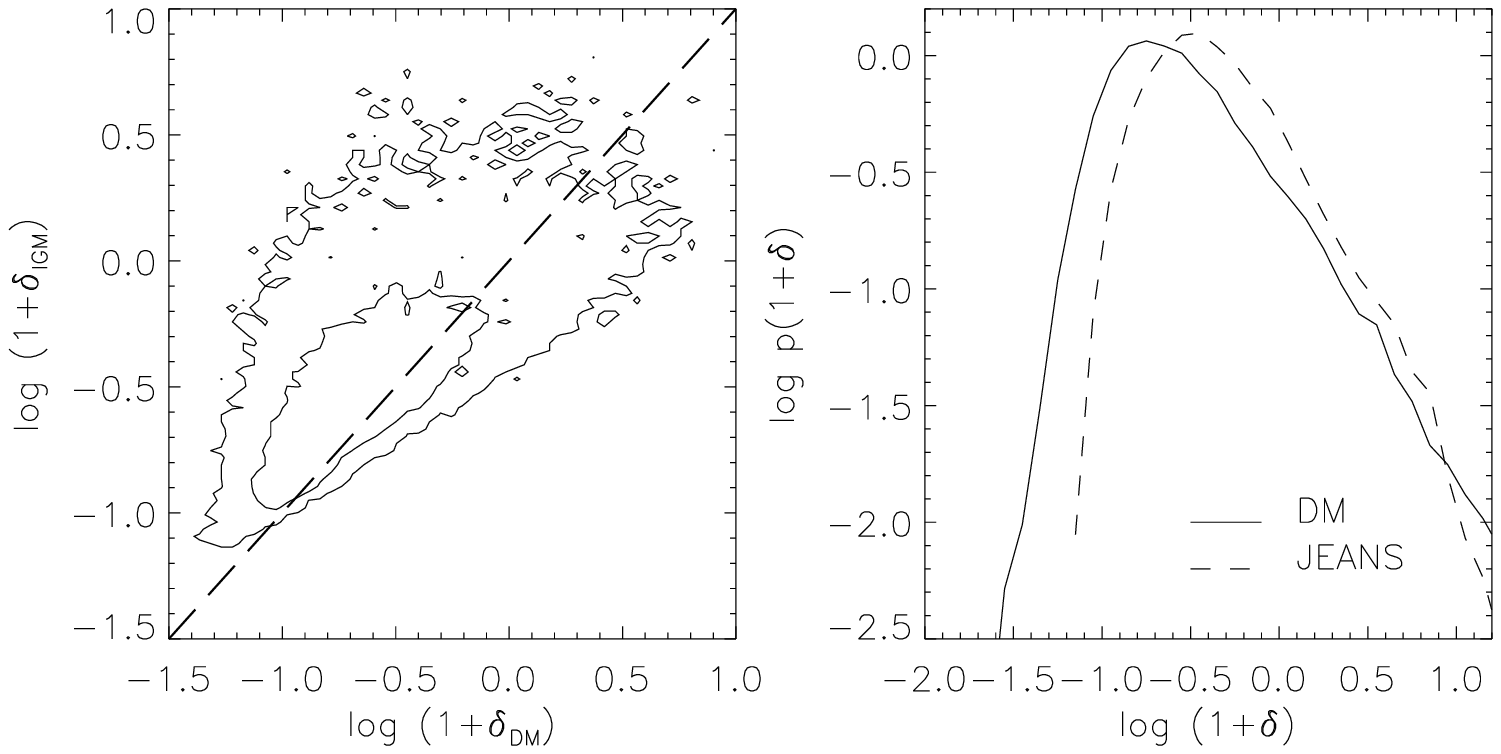}} 
\caption{Left panel: scatter plot of the dark matter density field smoothed on the 
Jeans length at $z=3$ and the dark matter density field; both these
fields are evaluated with SPH interpolation in $128^3$ mesh points.
$3\times 10^5$ points are shown; the 1 line is obtained by setting
$\delta_{IGM}=\delta_{DM}$. This plot has to be compared with Figure
\ref{fitscatter}, which shows the results from
hydro-simulations. Right panel: PDF of the dark matter (continuous
line) and PDF of the Jeans smoothed dark matter density field (dashed
line).}
\label{jeans}
\end{figure*}

Here we investigate the effect of smoothing the evolved rather than
the initial DM density field on a constant scale set by the Jeans
length at mean density. We use the DM distribution of the $z=3$ output
of the hydro-simulation to compute the dark matter density field 
on a cube of $128^3$ mesh points using SPH interpolation 
as described in the Appendix. The cell size of this mesh of 
$\approx 0.14$ co-moving Mpc approximately resolves the Jeans 
length $\lambda_J\sim 1$Mpc. We have
to choose a mesh because the filtering is done in 3D using fast Fourier
transforms. However, we note that also 1D smoothing along LOS is in
rough agreement with the 3D one. To model the Jeans  smoothing we convolve the DM density field
with a Gaussian filter $W=\exp(-k^2/2\,k_f^2)$, with $k_f=k_J\sim
7\mpc^{-1}$. By comparing Figure \ref{fitscatter} with Figure
\ref{jeans} one can see that the Jeans smoothed dark matter density
field is very different from IGM distribution in the hydro-simulation.
Given this large discrepancy between hydro-simulations and Jeans
smoothed dark-matter density field we choose not to analyse into the
details the density statistics as we did for the models of the previous subsection.

The main reason for the large difference between the true IGM density
distribution, and the one obtained from Jeans smoothing the DM density
field, is the simplification of a {\rm constant} smoothing length. In
reality, the Jeans length depends on temperature, and therefore should
be adaptive. Unfortunately, Fourier space filtering techniques smooth
all structures in the same way, since they are global operations. In
low-density regions, the amount of smoothing in hydrodynamical
simulations is small, but the Jeans smoothed density field is very
different from the original DM field. On the other hand, for large
over densities of the dark matter, Jeans filtering at the mean gas
density underestimates the amount of smoothing, the region at
$\delta_{IGM} < \delta_{DM}$, for large values of $\delta_{DM}$, is
more populated in Figure \ref{jeans} than in Figure \ref{fitscatter}.
As we will see  in the next section  the resulting flux statistics are
very different as well.

\section{A comparison of flux statistics for the  improved methods}
\label{fluxstat}
\begin{figure*}
\resizebox{0.8\textwidth}{!}{\includegraphics{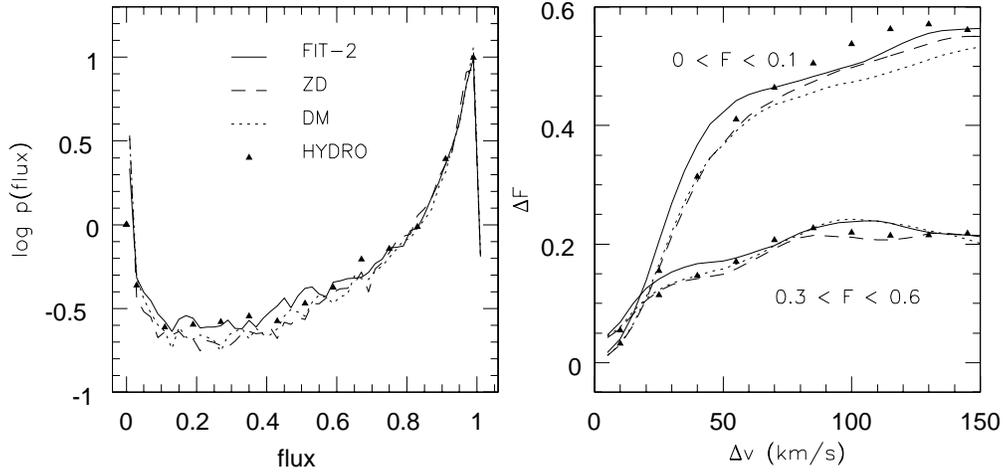}} 
\caption{PDF of the flux (left panel) and $\Delta F$ which, is related to the 
2-point PDF of the flux (right panel). Comparison between different
methods: FIT-2 (continuous line), ZD (dashed), DM (dotted). The
PDFs extracted from hydro simulations are represented by the triangles
(HYDRO). This plot has to be compared with Figure (\ref{flux}).}
\label{2PDFall}
\end{figure*}
In this section we compare the flux distribution 
obtained  with the different methods for modelling  the
gas density field. Figure \ref{2PDFall}
shows the 1-point probability distribution function for the flux
(left panel) and the mean flux difference $\Delta F$
(Eq.\ref{eq:deltaF}, right panel), for three models: {\it(i)} DM
(dotted line) is the PDF we get by setting $\delta_{IGM}=\delta_{DM}$;
{\it(ii)} ZD (dashed lines) is the PDF obtained with the `Zel'dovich
Displacement' method of Section \ref{ZEL} with a $k_f\sim 16
\mpc^{-1}$; {\it(iii)} FIT-2 (continuous line) is the PDF obtained with
the fitting technique of the previous Section and adding the scatter
inferred from the dark matter peculiar velocity field. We  also show 
the flux PDF of the hydro-dynamical simulation (triangles) which is 
the true PDF extracted from the $z=3$ output of the $\Lambda$CDM 
model.

All these methods try to predict the gas distribution 
from the dark matter distribution of the numerical simulation. 
We thereby assume that $v_{HI}\sim v_{IGM} \sim v_{DM}$ which we have
checked to be a good approximation. The  local abundance of 
neutral hydrogen is computed from the ionisation equilibrium equation,
eq. (\ref{ionization}). We further need to assume the temperature at 
mean density, $T_0$, and the slope of the temperature-density
relation, $\gamma$. We take  $T_0=10^{4.3}$K and $\gamma=1.2$. 
All simulated spectra have been  scaled to the same effective
optical depth, $\tau_{eff} \sim 0.27$ (which is a good fit to the observed
effective optical depth at redshift $z=3$).

The PDF of the flux is very similar for the three models. We performed
a quantitative comparison  based on the Kolmogorov-Smirnov (KS) test
which characterises the difference between two models from the maximum
absolute deviation, $d_{\rm KS}$, between the two cumulative flux
distributions (see Meiksin {\it et al.} 2001 for further details on KS
test applied to the PDF of the flux).  We calculate $d_{KS}$ for the
lognormal model (LOGN) and the `improved' model (PDF) obtained by
implementing the 1-point probability distribution function of the IGM
as well.  Fluxes obtained with TZA have not been computed as the large
scatter found (left panel, Figure \ref{compare}) suggests that this
approximation is the least accurate. The KS test was performed for  a
total of $3 \cdot 10^5$ pixels (Table \ref{tab:KS}).

The best agreement with the hydro-simulation is obtained with the 
FIT-2 method. The KS indicates a slightly better agreement than 
for the FIT-1 method. The ZD method gives better agreement 
than assuming that the gas traces the dark matter.   The 
PDF and LOGN methods are the least accurate; this means that 
the proposed fitting methods determine a better agreement in terms of
the pdf of the flux and can be considered as an effective improvement compared to
the PDF and LOGN methods.

\begin{table}
\caption{$d_{KS}$ values for the different methods proposed.}
\begin{center} 
\begin{tabular}{lrcr} 
\hline\hline 
Model & $d_{KS}$\\  
\hline\hline 
FIT-2 & 0.014  \\ 
FIT-1 & 0.018  \\
ZD  &  0.042   \\ 
DM  &  0.049   \\ 
PDF  &  0.102  \\ 
LOGN  & 0.147  \\ 
\hline\hline 
\label{tab:KS}
\end{tabular} 
\end{center} 
\end{table}

The methods based on the actual DM density distribution (FIT-1,
FIT-2, DM) also reproduce the 2-point flux distribution of the
hydro-simulation dramatically better than those based on linear theory
or the lognormal model as can be seen by comparing Fig. \ref{2PDFall}
(right panel) with Fig.~\ref{flux}. Note especially the significant
improvement in the shape of strong lines. As discussed above 
this implies that this statistic is mainly influenced by the 
correlations in the underlying dark matter density fields which 
are now the same for all the models.

\begin{figure*}
\resizebox{0.4\textwidth}{!}{\includegraphics{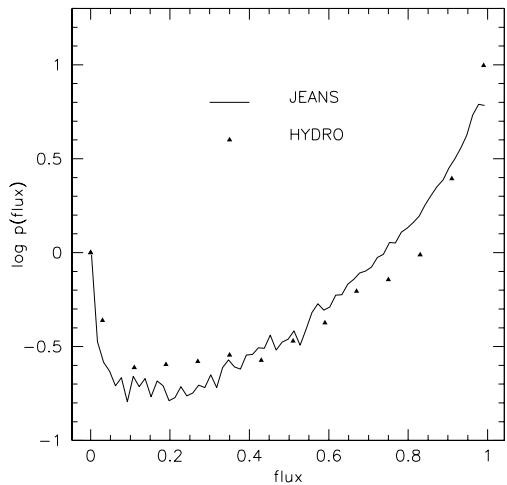}} 
\caption{PDF of the flux obtained from the Jeans smoothed 
dark-matter density field (continuos line). Filled triangles represent the hydro-simulation PDF.} 
\label{pdffinal}
\end{figure*}

We choose to make a final plot to quantify the differences between the
Jeans smoothed dark matter density field and the hydro-simulations in
terms of the PDF of the flux.
Results are shown in Figure \ref{pdffinal}.
The number of pixels per LOS is 128, significantly
lower than in the comparisons between the other improved  methods
\ref{imple}. To produce `Jeans smoothed' spectra we assume again 
that $v_{HI} \sim v_{DM}$ and a power-law equation of state for the dark matter
over density.  This time  there are also significant differences
between the the two models also in terms of the PDF of the flux.  The
hydro PDF predicts more dense regions than the Jeans smoothed one,
and correspondingly less regions with small densities. This is somewhat
counterintuitive, since the fitting technique smooths less in low
density regions compared with Jeans smoothing. However, we require both
types of spectra to have the same $\tau_{eff}$. Effectively, this means
that the two PDFs for $\delta$ are multiplied by a correction factor,
which can be seen as a scaling of the neutral hydrogen, in such a way
that the hydro PDF has a larger number of regions at large $\delta$
than the Jeans smoothed field. The $d_{KS}$ of the two distribution is
$\sim 0.1$ slightly better than the LOGN model value. 

The main conclusions of this Section are: {\it (i)} the use of 
the numerically simulated dark matter density distribution for the prediction
of the gas density distribution results in a significant improvement 
in the  comparison with the hydro-simulations; {\it (ii)} 
the modelling of the gas distribution using the fit 
to the mean relation between dark matter and gas density 
in the  hydro-simulations results in  much better
agreement than the methods where a  filtering scheme 
is applied to the initial or evolved density field.

\section{Discussion and conclusions}
\label{disc}
Many aspects of the warm photo-ionised Intergalactic medium can be well
modelled by hydrodynamical simulations. These are, however, still rather 
limited in dynamic range and lack the possibility  of extensive parameter 
studies due to limited computational resources. In order to overcome 
these problems we have tested several approximate methods for
simulating the Ly$\alpha$ forest in QSO absorption spectra. 
The modelling consists of two main steps (i) modelling the DM 
distribution and (ii) mapping  the DM distribution into a gas distribution. 

Methods which use an analytic description of the DM distribution 
like the lognormal model give a rather poor description of the 
gas distribution compared to numerical simulations. In Section 
\ref{secmet} we have shown that this results in  flux
PDFs which are in rough agreement with the PDF extracted from 
hydro-simulations. The agreement can be improved with  a variant of the 
lognormal model where a mapping between linear and non-linear IGM 
density fields calibrated by hydro-dynamical simulations is used. 
However, in both cases, the 2-point PDF of the flux differs
significantly from that obtained from numerical simulations. 
This is because the 2-point PDF is strongly affected by the 
underlying correlations in the DM density field, which are not 
well reproduced in models based on an extrapolation of linear theory. 
To take properly account of these correlations on a point-to-point 
basis numerical DM simulations are required. 

We thus tested a variety of schemes to relate the DM distribution 
of numerical simulations to the gas distribution. These  schemes are supposed 
to take into account that the gas is smoothed on a Jeans scale 
relative to the dark matter. We  first investigated  two
approximations based on the Zel'dovich approximation, the truncated 
Zel'dovich approximation 
(TZA) and a scheme  which we call Zel'dovich displacement (ZD). The
latter is based  on the assumption that the displacement between  DM 
and IGM at the same Lagrangian coordinate depends on the  
initial DM density field, filtered on a suitable scale. The TZA  reproduces 
the gas density field  very poorly in a LOS by LOS  comparison.
It is actually worse than if we assume that the  gas traces 
the DM faithfully. The ZD method which allows diffusion on a scale 
smaller than half of the Jeans length to mimic baryonic 
pressure fares somewhat better. The scatter in plots of 
predicted vs.  simulated densities is nevertheless only 
slightly smaller than in a model where gas  traces dark matter. 
We have also tested the ZD method with smoothing on  a global Jeans 
length and have again found poor agreement with spectra extracted 
from hydro-simulations. It may be that filtering techniques based on
the Zel'dovich approximation can reproduce some statistical properties 
of the Ly$\alpha$ forest like the column density distribution
reasonably well, but they fail in reproducing the flux distribution in
detail. 

To make progress we have thus investigated the relation of the  gas 
density and the DM density in the hydro-dynamical simulation in more
detail. The relation between $\delta_{IGM}$ and $\delta_{DM}$ can be well fit 
with a 3rd order polynomial.  There is considerable scatter around
the mean relation and we have examined the correlations between 
deviations from the fit with other physical quantities along
the LOS.  There is a weak correlation of these  deviations 
with  the filtered dark matter peculiar velocity gradient and a 
somewhat  stronger correlations with the gas temperature. 
This indicates that the deviations are due to  moderate or strong 
shocks in the gas component. 

Combining the DM simulations with the fitted relation between 
DM density and gas density gives good results for both the one
and two-point distribution of the flux.  If we introduce 
an  additional correlation of the gas density with the DM peculiar
velocity gradient as found in the hydro simulations the agreement 
is further improved (a method which we called FIT-2).  These fitting 
methods give significantly better results than the other methods 
we have discussed.  
 
Smoothing of the dark matter density field with a constant global
Jeans scale calculated  for mean density and temperature  results in a gas 
distribution  very different from that found in hydro-simulations.  
This is not too astonishing  as it does not take into account that 
the Jeans length depends on temperature and density. Jeans smoothing 
at the mean temperature  overestimates the  smoothing in low density 
regions and underestimates it at higher  density. This leads  to significant 
differences of the flux  statistics compared to hydro-simulations.  
Smoothing of the ${\it evolved}$ dark matter density field on a 
global Jeans scale  is therefore not a promising technique. 
The reason  for the success of our FIT-1 and FIT-2 schemes is their 
adaptive nature which  takes  into account -- at least to some extent
--  the density/temperature dependence of the Jeans scale. 

We conclude that large high-resolution DM simulations combined with 
a two-parameter fit of the DM density gas density relation obtained 
from hydro-dynamical simulations are the best compromise between 
computational expense  and accuracy when a large dynamic range and/or 
an extensive parameter study are required.

\section*{Acknowledgments}
We thank Bepi Tormen, Francesco Miniati, Lauro Moscardini, Joop Schaye
and Simon White for useful discussions and technical help.  MV
acknowledges partial financial support from an EARA Marie Curie
Fellowship under contract HPMT-CT-2000-00132. TT thanks PPARC for the
award of an Advanced Fellowship. This research was conducted in
collaboration with Cray/SGI, utilising the COSMOS super computer at the
Department for Applied Mathematics and Theoretical Physics in
Cambridge. This work was supported by the European Community Research
and Training Network `The Physics of the Intergalactic Medium'.

\appendix
\section{SPH interpolation along LOS}
\label{SPH}
In this Appendix we describe the SPH computation of physical
quantities along a given line of sight through the box. We follow the
same procedure described in (Theuns {\it et al.} 1998). We divide the
sight line into $N\sim 2^{10}$ bins of width $\Delta$ in distance $x$
along the sight line. For a bin $j$ at position $x(j)$ we compute the
density and the density weighted temperature and velocity for the gas
and density and weighted velocity for the dark matter from:
\begin{eqnarray}
\rho_X(j)   &=& \sum_i {\cal W}_{ij}\\ 
(\rho v)_X(j) &=& \sum_i v_{X(i)} {\cal W}_{ij}\\ 
(\rho T)_X(j) &=& \sum_i T(i) {\cal W}_{ij}\, . 
\end{eqnarray}
$X(i)$ is a label indicating the abundance of species $X$ of particle
$i$ ($X=\H$; $X=$IGM and $X=$DM denotes neutral hydrogen, total gas
and dark matter density, respectively).

Here, ${\cal W}_{ij} = m W(q_{ij})/h_{i}^3$ and $m$ is the SPH
particle mass which is the same for all SPH particles (but different
for DM and gas particles). For $W$ we use the M4 spline (Monaghan 1992)
given by
\begin{eqnarray}
W(q) &=& {1\over \pi} (1.+q^2\,(-1.5+0.75q)) \,\,\,{\rm if}\,q\le 1\nonumber\\
     &=& {1\over \pi} (0.25\,(2.-q)^3) \,\,\,{\rm if}\,1\le q\le 2\nonumber\\
     &=& 0 \,\,\,{\rm elsewhere.}
\label{eq:kernel}
\end{eqnarray}
We have defined: \be q_{ij} = {|\bfx(i)-\bfx(j)|\over h_{i}}, \ee where
$\bfx(i)$ and $h_{i}$ are the position and SPH-smoothing length of
particle i.  Note that $h$ is defined in such a way that on average 32
particles are within $2h(i)$ from particle i.  In this way, for each
pixel along the LOS, we compute the contribution of all the particles
which influence this region with a weight given by
eq. (\ref{eq:kernel}); this is the `scatter' interpretation (see, for
example, Hernquist \& Katz 1989). For the computation of the spectra we
label bins according to velocity and we adopt the procedure described
in Section \ref{secmet}.

\end{document}